\documentclass[twocolumn,prb,showpacs,preprintnumbers,amsmath,amssymb,floatfix]{revtex4}

\usepackage{epsfig,psfrag}
\usepackage{dcolumn}
\usepackage{bm}
\usepackage{graphicx}
\usepackage{color}

\begin{document}

\title{Wavevector-dependent spin filtering and
spin transport through magnetic barriers in
graphene}
\author{L. Dell'Anna$^1$ and A. De Martino$^2$}
\affiliation{
$^1$International School for Advanced Studies (SISSA), 
Via Beirut 2-4, I-34014 Trieste, Italy\\
$^2$Institut f\"ur Theoretische Physik, Universit\"at zu K\"oln,
Z\"ulpicher Stra\ss e 77, D-50937 K\"oln, Germany
}

\date{\today}

\begin{abstract}
We study the spin-resolved transport through magnetic nanostructures 
in monolayer and bilayer graphene. We take into account both the orbital 
effect of the inhomogeneous perpendicular magnetic field as well as 
the in-plane spin splitting due to the Zeeman interaction and to
the exchange coupling possibly induced by the proximity of a 
ferromagnetic insulator. We find that a single barrier exhibits a
wavevector-dependent spin filtering effect at energies close 
to the transmission threshold. This effect is significantly
enhanced in a resonant double barrier configuration, where 
the spin polarization of the outgoing current can be increased 
up to 100\% by increasing the distance between the barriers. 
\end{abstract}


\maketitle

\section{Introduction}
\label{intro}

The electronic properties of graphene\cite{geim,kim}
have attracted in the last four years huge experimental as well
as theoretical attention.\cite{reviews} Besides its fundamental 
interest as a new type of two-dimensional electron liquid,
graphene is in fact regarded as a promising material for 
future nanoelectronic devices,\cite{reviews} 
in particular in the field of spintronics,\cite{spintronics} 
due to its small intrinsic spin-orbit\cite{so1,so2,so3} 
and hyperfine interactions.\cite{bjorn} 
Indeed several recent 
experiments\cite{expspin1,expspin2,expspin3,expspin4} 
have by now demonstrated spin injection and detection in a
single layer of graphene sandwiched between ferromagnetic
metal electrodes and observed coherent spin transport 
over micrometer scale distances.

Motivated by these developments, in this paper we 
focus on the problem of spin resolved transport 
through {\em magnetic nanostructures} in graphene. 
The studies of graphene's transport properties 
through magnetic barriers and more complex structures
\cite{ale,sim,cserti,tarun,wolfgang,zhai,tahir,masir0,masir1,masir2,
heinzel,kormanios,ghosh,luca}
address the problem of controlling the confinement 
and the transport of charge carriers by means of 
appropriate configurations of an 
external magnetic field inhomogeneous 
on sub-micron scales. Different types of 
magnetic  nanostructures have been envisioned and their
properties investigated, e.g., barriers 
\cite{ale,masir0,masir1,masir2}, dots \cite{ale}, wires  \cite{tarun} 
and superlattices. \cite{luca} 

In all these works the spin degree of freedom 
has been completely neglected. This is justified because
of the smallness of both Zeeman splitting and spin-orbit 
coupling\cite{so1,so2,so3} in graphene. However, the continuous 
and rapid improvements in sample preparation and experimental technology 
and resolution in graphene research call for a refinement of the 
theoretical analysis to incorporate such finer effects. 
More importantly, it has recently been argued that 
local ferromagnetic correlations can be induced in 
graphene by several different mechanisms, e.g., proximity of
a ferromagnetic insulator,\cite{exchange1,exchange2,exchangebil}
Coulomb interactions,\cite{exchangecoul} presence of 
defects,\cite{exchangedefect} application of an
electric field in the transverse  direction in 
nanoribbons.\cite{exchangeribbons}
The ferromagnetism leads to a spin splitting effectively similar to
a Zeeman interaction but of much larger magnitude.
For example, a ferromagnetic
insulator deposited on top of a graphene layer 
has been predicted to produce a spin splitting of 
up to $5$\, meV. \cite{exchange1} 
This is comparable with the orbital 
energy in a field of $1$ T, which is of order of $25\,$ meV, 
and thus may have important effects.
The spin transport through ferromagnetic
graphene has already received some 
attention,\cite{exchange1,exchange2,yokoyama,niu,namdo,li} 
but to the best of our knowledge all works so far have focussed 
only on spin effects and did not consider orbital effects. 
Aim of this paper then is to fill the gap and investigate 
this problem by fully taking into account the quasiparticle's 
spin dynamics as well as its orbital motion in an inhomogeneous 
magnetic field. 

In this paper we shall focus on the effects of  
Zeeman and exchange spin splitting and postpone the interesting 
problem of spin-orbit coupling to a future work.
We show that the spin-resolved transmissions 
exhibit  a strong dependence on the incidence angle, 
which allows in principle for a selective transmission 
of spin-up and spin-down electrons. This effect can be 
qualitatively understood by a simple semiclassical argument. 
The bending of the electron trajectories under the 
barrier depends on the energy and thus, in the presence
of spin splitting, is different for the two spin projections.
The magnitude of the polarization that can be achieved
depends on the spin splitting and can be
very large in the presence of a large splitting,
as, e.g., that originating from a proximity-induced
exchange field.
Moreover, in a resonant double barrier configuration, the 
polarization can be enhanced by 
increasing the distance between the barriers.
In this case, in fact, even for relatively small 
splitting there exists an energy range where
the polarization reaches values close to one
for large enough distance.

While there exist several experimental techniques to produce
magnetic barriers, for concreteness we shall have in mind 
the case of the fringe fields created by a ferromagnetic stripe 
deposited on top of the graphene sample. The magnetic field 
generated by the stripe is known analytically. \cite{ibrahim}
The corresponding vector potential in the Landau gauge
can be written as ${\bf A}=A(x,z)\hat y$, where
\begin{equation}
 A(x,z) = \left[
\int_{-\infty}^x B_z(x',0) dx' - \int_0^z B_x(x,z') dz'
\right]  ,
\label{vecpot}
\end{equation}
and we assume that the system is translationally invariant 
in the $y$ direction. We shall neglect possible corrugations of 
graphene and assume that it lies flat in the $z=0$ plane. 
Then the orbital dynamics is only affected by the perpendicular 
component $B_z$, while both $B_x$ and $B_z$ contribute to 
the Zeeman interaction.  In addition we shall also include in the 
Zeeman term the effects of 
a possible exchange field. \cite{exchange1,exchange2}

The transport properties of such structures 
can be calculated by means of the Landauer-B\"{u}ttiker 
formalism in terms of  the spin-resolved
transmission matrix 
$t_{\sigma\sigma'}$, 
which gives the probability amplitude for a quasiparticle incident 
on the magnetic structure from the left 
with spin projection $\sigma'$ to be transmitted 
with spin $\sigma$. \cite{zhai2005b,zhai2006,nikolic,ratchet}
The matrix $t_{\sigma\sigma'}$  
depends on particle's energy $E$ and incidence angle $\phi$ (see below)
and must satisfy certain general symmetry requirements.\cite{zhai2005b}
We shall be interested in the spin-resolved conductances
\begin{equation}
\label{G}
G_{\sigma \sigma'}=G_0 \int^{\pi/2}_{-\pi/2} d\phi \, \cos \phi
\, | t_{\sigma\sigma'} |^2, 
\end{equation}
and, with total conductance $G=\sum_{\sigma \sigma'} G_{\sigma\sigma'}$
and spin quantization axis along the $x$ direction,
the polarization vector of transmitted current 
for  a spin-unpolarized incident current \cite{nikolic}
\begin{eqnarray}
\label{Px}
&&{\cal P}_x =  \frac{G_0}{G}  
\,  \sum_{\sigma \sigma'} \sigma 
|t_{\sigma\sigma'} |^2 ,\\
\label{Py}
&&{\cal P}_y = \frac{2G_0}{G}  
\, \Re \sum_\sigma
t^{}_{\uparrow\sigma} 
t^*_{\downarrow\sigma} , \\
\label{Pz}
&&{\cal P}_z = \frac{2G_0}{G}
\,  \Im \sum_\sigma
t^*_{\uparrow\sigma} 
t^{}_{\downarrow\sigma},
\end{eqnarray}
with $G_0=\frac{2e^2}{h} \frac{EL_y}{2\pi\hbar v_F}$ 
and their integrated value 
\begin{equation}
\label{P_int}
P_i=\int^{\pi/2}_{-\pi/2} d\phi \, \cos \phi {\cal P}_i .
\end{equation}

The rest of this paper is organized as follows. 
In Sec.~\ref{singlelayer} we discuss the spin-dependent 
transmission and transport  through rectangular and double 
resonant barriers in single-layer graphene. In Sec.~\ref{secbilayer} 
we address the same problem in bilayer graphene and finally in 
Sec.~\ref{conclusions} we draw our conclusions. 

\section{Single-layer graphene}
\label{singlelayer}

Let us then start focussing on single-layer graphene. 
We shall neglect disorder and interaction effects and focus 
on a single  $K$ point, where ballistic motion
of charge carriers in an external 
magnetic field ${\bf B}=B_x \hat x + B_z \hat z$, 
is described by the Dirac-Weyl (DW) Hamiltonian 
\begin{eqnarray}
&&H = v_F {\bm \tau} \cdot \left( {\bf p} +\frac{e}{c} {\bf A}\right) + 
H_{spin},\label{DW}\\
&&H_{spin} = 
\frac{g_s\mu_B}{2} { \bf B}^{eff}  
\cdot {\bm \sigma}, \label{hspin}
\end{eqnarray}
where ${\bf p}=p_x \hat x + p_y \hat y$, $v_F\approx 10^6$ m/s
is the Fermi velocity,  $\mu_B$
the Bohr magneton 
and $g_s\approx 2$ the effective Land\'e factor. 
$H_{spin}$
is the sum of the Zeeman interaction due to 
the magnetic field and, possibly, the 
proximity-induced exchange splitting, 
with an estimated value\cite{exchange1}
$h_{ex} \approx 5$ meV 
(corresponding to a Zeeman interaction 
with a field of about $86$ T).
The vector of Pauli matrices $\bm \tau=\tau_x \hat x +\tau_y \hat y$ 
(resp. $\bm \sigma =\sigma_x \hat x + \sigma_y \hat y+ \sigma_z \hat z$) 
acts in sublattice space (resp. spin space), and
the wavefunction $\Psi$ is a four component  object, 
$\Psi^T=(\Psi_{A\uparrow}, \Psi_{B\uparrow}, \Psi_{A\downarrow}, 
\Psi_{B\downarrow}$) (the superscript  $T$ denotes transposition).
 ${\bf A} = A(x,z) \hat y$ is the vector potential  
in the Landau gauge, with 
$A(x,z)$ given in Eq. (\ref{vecpot}).
Since the Hamiltonian is translationally invariant in the $y$ 
direction and $p_y$ is conserved, the wavefunction can be 
written as $\Psi = e^{ik_y y}\psi $ and the DW equation 
reduces to a one-dimensional problem.
In the following we use rescaled quantities: 
$x\rightarrow x \ell_B$,
$k_{x,y}\rightarrow k_{x,y} \ell_B^{-1}$,
$A\rightarrow A B\ell_B$ and 
$E\rightarrow E\frac{\hbar v_F}{\ell_B}$, where
$B$ is a typical value of magnetic field in the problem.
The magnetic length is 
$\ell_B=\left(\frac{\hbar c}{eB}\right)^{1/2}
\approx 26/\sqrt{B[T]}$ nm and the orbital 
energy scale is 
$E_m=\frac{\hbar v_F}{\ell_B} \approx 25\sqrt{B[T]}$ meV,
where $B[T]$ is the magnetic field strength expressed in Tesla.
\begin{figure}
{\includegraphics[width=8.5cm]{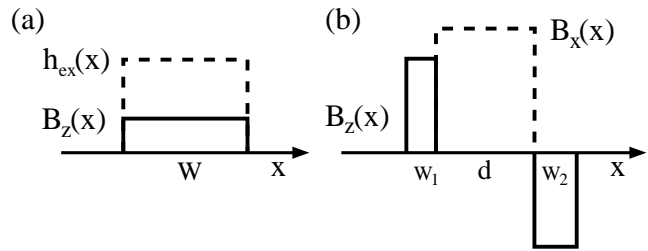}}
\caption{Profiles of magnetic field and
exchange field considered in this paper.
In case (b)
the exchange field is included in $B_x$.}
\label{profile}
\end{figure}

\subsection{Rectangular barrier}

First we discuss the tunneling of a spinful DW quasiparticle 
through a rectangular magnetic barrier of width $W$ 
with an effective Zeeman field
\begin{equation}
{\bf B}^{eff} =
\lambda \, \theta(W/2-|x|) \hat z,
\end{equation}
see Fig.~\ref{profile}(a). The value of $\lambda$ 
ranges from $0.002$ for a Zeeman field of $1$ T up to 
$\lambda\approx 0.2$ if an
exchange splitting of $5$ meV is included. 
With spin quantization axis along  $\hat z$,
the Hamiltonian (\ref{DW}) is diagonal in spin space 
and the two spin components can be treated separately.
Accordingly, the wavefunctions are just two-component
spinors. The solution to Eq. (\ref{DW}) with energy $E$  
(assumed positive for definiteness)
and spin projection $\sigma$ describing a scattering state
incoming from the left
can be written in the left and right "leads" 
(i.e. the non-magnetic regions $x<-W/2$ and $x>W/2$)
as
\begin{eqnarray}
\psi_{\sigma}(x<-W/2)=\frac{1}{\sqrt{k^i_x}}{\cal W}_0(x) 
\left( \begin{array}{c}
1\\r_{\sigma\sigma}
\end{array} \right),\\
\psi_{\sigma}(x>W/2)=\frac{1}{\sqrt{k^f_x}}{\cal W}_0(x) 
\left( \begin{array}{c}
t_{\sigma\sigma}\\0
\end{array} \right), 
\end{eqnarray}
where $r_{\sigma\sigma'}$ (resp. $t_{\sigma\sigma'}$)
is the probability amplitude for a quasiparticle incident  from the left 
with spin projection $\sigma'$ to be reflected (resp. transmitted)
with spin $\sigma$. The matrix ${\cal W}_0$ is given by
\begin{equation}
{\cal W}_0(x) = \left(\begin{array}{cc}
1 & 1 \\
e^{i\phi(x)} & - e^{-i\phi(x)}
\end{array} \right) e^{i \tau_z k_x(x)x}
\end{equation}
with $k_x(x)=\sqrt{E^2-[k_y+A(x)]^2}$ and 
$\pm e^{\pm i\phi(x)} = \frac{\pm k_{x}(x)+i[k_y+A(x)]}{E}$.
In the leads we use the parameterization:
\begin{eqnarray}
&&k_y= E \sin \phi_i = E \sin \phi_f  - \Phi,\\
&&k_x^i = k_x(x<-W/2) = E \cos \phi_i ,\\ 
&&k_x^f= k_x(x>W/2) = E \cos\phi_f,
\end{eqnarray}
where $\Phi = A(x>W/2)$ is the total perpendicular magnetic flux 
through the barrier per unit length in the $y$ direction.
For $\Phi>0$ the emergence angle $\phi_f$ is larger than the
incidence angle $\phi_i$. Thus a finite transmission is only 
possible if $\phi_i$ is smaller than the critical angle 
$\phi_c=\arcsin (1-\Phi/E)$. For
energy smaller than the threshold value 
$E_{th}=|\Phi|/2$ the transmission vanishes 
for any incidence angle.\cite{ale,luca} 
This condition has a simple geometrical
interpretation. 
In momentum space the dispersion cone 
after the barrier is shifted with respect to the
cone in the region before the barrier
by $-\Phi$ along the $k_y$ axis.
Thus if the radius of the fixed energy circle (the Fermi line)
is smaller than $\Phi/2$, the equal energy sections
of the two cones do not overlap, 
implying that $k_x^f$ is imaginary for any $k_y$,
i.e., the barrier is perfectly reflecting.
Note that neither the emergence angle $\phi_f$
nor the critical angle $\phi_c$ depend on spin.  
\begin{figure}
{\includegraphics[width=8cm]{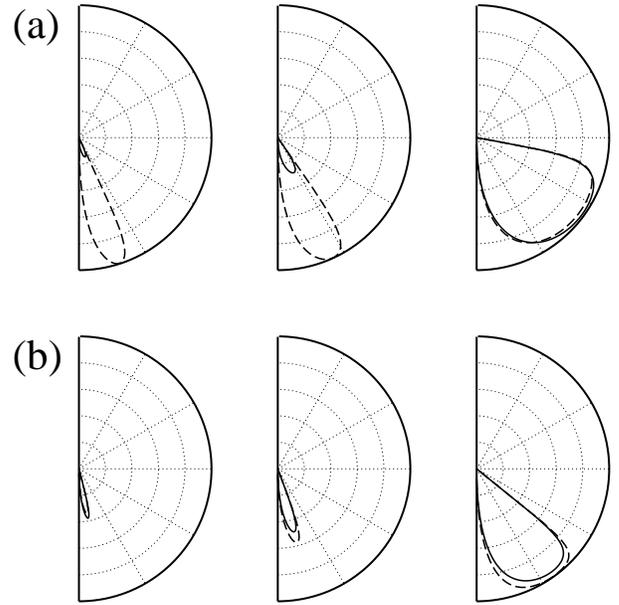}}
\caption{(a) Angular plot of the
transmission through a rectangular magnetic 
barrier of width $W=2$ (in units of $\ell_B$)
with $\lambda =0.2$ for spin-up (solid lines) and 
spin-down (dashed lines) quasiparticles at energy  
$E=1.05$, $1.1$, $1.7$ (from left to right, in units of $E_m$);
(b) the same for a barrier of width
$W=6$ at energy $E=3.05$, $3.1$, $3.7$.
Here we set $B=1$ T, so that $E_m\approx 25$ meV and
an energy increment of $0.1$ corresponds to $2.5$ meV.
}
\label{figrecbar}
\end{figure}
In the barrier region $|x|<W/2$ the spin-$\sigma$ wavefunction 
can be written as\cite{ale} 
\begin{eqnarray}
&&\psi_\sigma(x)= {\cal W}^\sigma_{B}(x)\left( \begin{array}{c}
a\\ 
b
\end{array}\right),\\
&&{\cal W}^\sigma_B(x) = \left(\begin{array}{cc}
D_{p_\sigma}(q) & D_{p_\sigma}(-q) \\
\frac{i\sqrt{2}}{E_\sigma}D_{p_\sigma+1}(q) & 
\frac{-i\sqrt{2}}{E_\sigma}D_{p_\sigma+1}(-q) 
\end{array} \right),
\end{eqnarray}
where $D_p(z)$ is the parabolic cylinder function, \cite{gradshteyn}
$q=\sqrt{2}\left[A(x) + k_y\right]$,
$p_\sigma=\frac{E^2_\sigma}{2}-1$, 
$E_\sigma=E- \sigma \lambda  $, 
and $a$ and $b$ complex amplitudes.
Continuity of the wavefunction 
then leads to the matching condition across the magnetic barrier:
\begin{equation} 
\left( \begin{array}{c}
1 \\ r_{\sigma\sigma} \end{array}\right)= \left( \frac{\cos 
\phi_i}{\cos \phi_f}\right)^{1/2}
\hat T \left( \begin{array}{c} 
t_{\sigma\sigma} \\ 0 \end{array}\right),
\label{matching}
\end{equation}
where the (spin dependent) transfer matrix $\hat T$ is given by
\begin{equation}
\hat T = \left[{\cal W}_0(-W/2)\right]^{-1} 
 {\cal W}^\sigma_B(-W/2) 
\left[{\cal W}^\sigma_B (W/2)\right]^{-1}
{\cal W}_0 (W/2).
\nonumber
\end{equation}
The transmission coefficient can directly be read 
from Eq. (\ref{matching}):
\begin{equation}
|t_{\sigma\sigma}|^2 = \frac{\cos \phi_f}{\cos \phi_i} 
\frac{1}{\left| T_{11} \right|^2}.
\end{equation}

The spin-resolved transmission
as a function of the incidence angle 
is illustrated in Fig.~\ref{figrecbar}
for two different barrier widths.
Remarkably we find that, even though the
critical angle $\phi_c$ is the same 
for both spin projections, there is a large 
difference between the spin-up and spin-down 
transmissions within an energy range $\lambda$
from the common threshold energy $E_{th}$, see
Fig.~\ref{figrecbar}(a). 
This is due to the very sharp angular and
energy dependences of the transmission onset.
However, this effect becomes very small as soon as
$\lambda/E_{th} \ll 1$, as one can see in 
Fig.~\ref{figrecbar}(b), where $E_{th}=3$.

Before moving to more complex structures, 
it is interesting to briefly consider the limit
$B\rightarrow \infty$, $W\rightarrow 0$ with fixed $BW=\Phi$,
where the magnetic field profile reduces to a $\delta$ function,
$B(x)= \Phi \delta(x)$.
Using the asymptotic behavior
of the parabolic cylinder function $D_p(z)$
for $p\rightarrow \infty$ and $z\rightarrow 0$ with 
$z\sqrt{p}$ finite: \cite{watson}
\begin{equation}
D_{-p-1}(\pm iz)= \frac{\sqrt{\pi}(p/e)^{p/2}}{\Gamma(1+p)} 
e^{\mp i z\sqrt{p}} \left( 1+ O(p^{-1/2})\right),  \nonumber
\end{equation} 
after some lengthy algebra we obtain the compact result
\begin{equation}
{\cal W}^\sigma_B(-W/2)\left[ {\cal W}_B^\sigma(W/2) \right]^{-1}
 \rightarrow e^{i\sigma \lambda_z \tau_x},
\label{matchingdelta}
\end{equation}
where the dimensionless Zeeman coupling is now given by
$\lambda_z =  \frac{\mu_B \Phi}{\hbar v_F}$. 
An elementary calculation then obtains the transmission as 
\begin{equation}
\left|t_{\sigma\sigma}\right|^2=
\frac{\cos \phi_i \cos \phi_f}{\cos^2 \lambda_z \cos^2 
\frac{\phi_i +\phi_f}{2}+
\sin^2 \lambda_z \cos^2 \frac{\phi_i-\phi_f}{2}}, 
\label{trandelta1}
\end{equation}
which, in contrast to the case of a barrier of finite width, 
is spin independent.
We note that if one solves the scattering problem 
by considering the Hamiltonian (\ref{DW}) directly with 
${\bf B}^{eff}=BW\delta(x)\hat z$
and imposing the matching condition obtained by integrating (\ref{DW}) 
across the origin with the prescription $\delta(x)\theta(x)=\delta(x)/2$, 
one obtains Eq. (\ref{trandelta1}) with $\lambda_z$ replaced by 
$\tilde \lambda_z=2\arctan \frac{\lambda_z}{2}$.
The precise functional dependence of the transmission 
on the $\delta$-function strength depends in fact
on the regularization, 
similarly to the case of an electrostatic barrier. \cite{kellar}
Since the use of the prescription $\delta(x)\theta(x)=\delta(x)/2$ 
in the DW first-order differential equation has been 
criticized,\cite{kellar} in the rest of the paper 
we shall use Eq. (\ref{matchingdelta}).

\subsection{The resonant double barrier}
\label{resonantbarrier}

\begin{figure}
{\includegraphics[width=8cm]{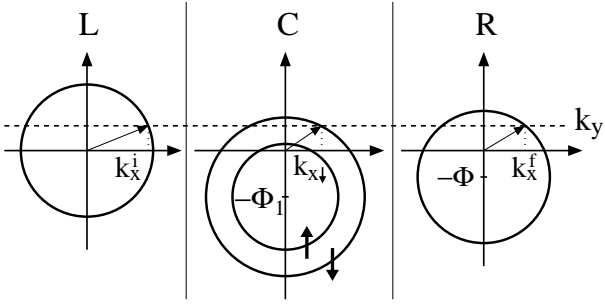}}
\caption{Illustration of the kinematics
of the transmission through a resonant 
double magnetic barrier.  
The solid circles represent the 
constant energy contours (Fermi lines)
before (resp. after) the structure (L, resp. R)
or in between the barriers (C).
In the leads L and R the circles are doubly spin degenerate. 
In C the vector potential horizontally shifts the cones along 
the $k_y$ axis, so that they are centered at 
$k_x=0$, $k_y=-\Phi_1$. The spin splitting in C
vertically shifts the cones upwards (resp. downwards)
for spin up (resp. spin down). The horizontal dashed
line represents the fixed $y$ component of the momentum, 
conserved across the barriers. $\Phi=\Phi_1+\Phi_2$ is the 
total flux through the structure (per unit length in 
the $y$ direction).
It is clear from 
the picture that if $|E-\lambda_x|<|k_y+\Phi|<|E+\lambda_x|$
a spin-up particle propagates in C via an evanescent wave. 
}
\label{geometry}
\end{figure}
\begin{figure}
{\includegraphics[width=8cm]{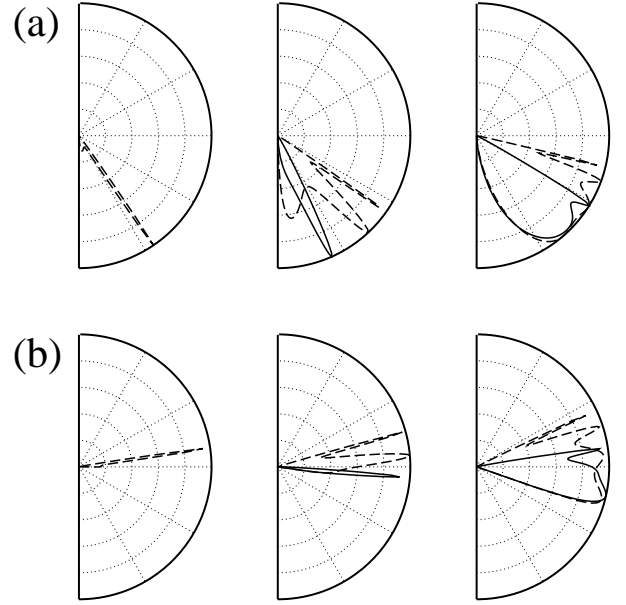}}
\caption{Angular plot of the transmission for 
the profile in Fig.~\ref{profile}(b) with
$d=6$, $\lambda_{x}=0.2$, $\lambda_z=0$.
Solid lines are for spin-up and dashed lines for spin-down. 
(a) Barriers of equal width $W_1=W_2=2$ and energy 
$E=1.02,\, 1.18,\, 1.5$ (from left to right);
(b) the same as in (a) but for barriers of different 
width ($W_1=1$ and $W_2=2$).
}
\label{Tdoublerecbar}
\end{figure}

We now discuss the case of a resonant structure
consisting of two rectangular magnetic barriers with 
opposite signs of the magnetic field and non-vanishing
in-plane Zeeman splitting in between, as illustrated 
in Fig.~\ref{profile}(b):
\begin{eqnarray}
&&{\bf B}= \left\{
\begin{array}{rr} 
B\hat z, & 0<x<W_1 \\
B_{x} \hat x, & W_1<x<W_1+d \\
-B\hat z, & W_1<x-d<W_1+W_2\\
0, & \text{otherwise}
\end{array}\right. 
\end{eqnarray}
This profile should qualitatively model the realistic 
configuration of the stray field produced by a ferromagnetic 
stripe, which, in addition to the normal component, also 
contains an in-plane component $B_x$. 
Inclusion of this component is crucial for the proper
treatment of the spin dynamics.\cite{zhai2006}

First, we neglect the Zeeman term under the barriers,
so that the problem is again diagonal in spin, with spin quantization
axis along the $x$ direction. By way of a simple geometric argument, 
illustrated in Fig.~\ref{geometry}, we argue that this structure 
exhibits a strong wavevector-dependent spin filtering effect. 
Indeed, in the region C between the barriers  the dispersion cones 
for spin-up and spin-down particles are equally shifted by 
$-\Phi_1=-BW_1$ along the $k_y$ axis with respect to the 
cones in the left (L) lead. Moreover the cones
are also shifted (say for $B_x>0$) 
upwards (resp. downwards) by the in-plane Zeeman splitting,
so that the radius (the Fermi momentum) increases (resp. decreases)
by $\lambda_x=\mu_B B_x/E_m$. 
As a result there exists a range of incidence angles in which 
the spin-down modes
propagate via travelling waves in the central region, whereas 
the spin-up modes 
only exist as evanescent waves and their transmission through
the structure is exponentially suppressed with the distance 
between the barriers. 
Formally, this can easily be seen from
the expression of the $x$ component of the momentum 
in the region C, $k_{x\sigma} = \sqrt{(E-\sigma \lambda_x)^2-(k_y+\Phi_1)^2}$, 
which is real for $\sigma=\downarrow=-$ and pure imaginary 
for $\sigma=\uparrow=+$
as long as $|E-\lambda_x|<|k_y+\Phi | < |E+\lambda_x|$. 
If $|E-\lambda_x|>\Phi/2$ the transmission 
$|t_{\uparrow\uparrow}|^2$ at any incidence angle is 
fully suppressed for large enough $d$
and spin-up modes do not significantly contribute to 
the transport through the structure.

The exact calculation of the transmission coefficients
follows the same lines as for the single rectangular barrier
discussed in the previous section. The results are 
illustrated in Fig.~\ref{Tdoublerecbar}.
Fig.~\ref{Tdoublerecbar}(a) shows indeed that in a certain range
of incidence angles  the transmission for spin-up particles
practically vanishes. By changing the ratio of barrier widths 
$W_1/W_2$ one can also control the position of the center 
of this interval, as illustrated in Fig.~\ref{Tdoublerecbar}(b).
The width of this range clearly depends on $\lambda_x$.
It is then crucial to have a large in-plane spin splitting 
if one is to observe this effect.

A simple closed formula for the transmission is easily 
obtained in the limit of $\delta$ barriers of equal 
and opposite strengths $\Phi_1=-\Phi_2=\Phi$:
\begin{equation}
\left| t_{\sigma\sigma}\right|^2=
 \frac{\left( \cos k_{x\sigma}d \right)^{-2}}
{1 + 
\frac{ \left[ k_y(k_y+\Phi)  - E(E-\sigma \lambda_{x})\right]^2}
{k_x^2 k_{x\sigma}^2} 
\tan^2 k_{x\sigma}d },
\label{transupdown}
\end{equation}
which explicitly exhibits the features discussed above and 
reproduces quite well the transmission for the case of double
rectangular barriers. 
Resonances occur at  $k_{x\sigma}d=\pi n$, 
with $n$ a positive integer, where  $|t_{\sigma\sigma}|^2=1$. 
Upon increasing $d$, the number of resonances increases and they also 
become narrower. Interestingly, the positions  of the resonances are 
different for spin-up and spin-down electrons.

Next we consider the general situation where we do not
neglect the Zeeman splitting under the barriers. 
Then with spin quantization axis along the $x$ direction, 
spin-flips can take place at the barriers, hence,
in contrast to the previous case, an incident particle 
upon transmission or reflection can change its spin state 
and the spin-filtering effect be spoiled. As we will see, however,
the effect still survives close to some thresholds.
\begin{figure}
 {\includegraphics[width=9cm]{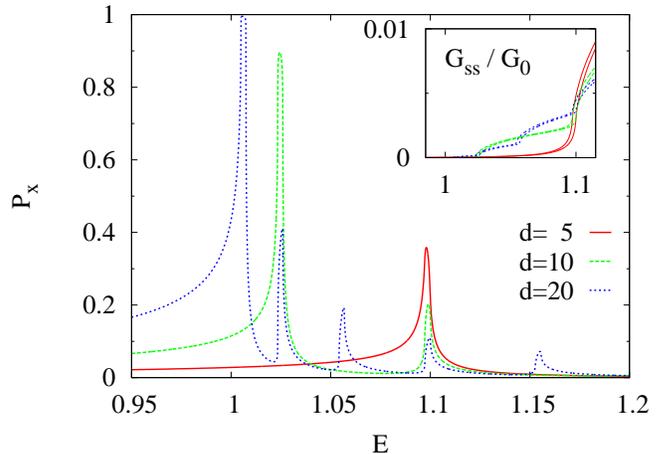}}
\caption{(Color online)
Plot of the $x$ component of the polarization vector, 
$P_x$, and, 
in the inset, the spin-resolved conductances $G_{\uparrow\uparrow}$ 
and $G_{\downarrow\downarrow}$ as functions of $E$,  
for a structure consisting of two $\delta$ barriers 
with opposite signs of the magnetic field at distance 
$d=5$ (solid red line), $10$ (long-dashed green line),
$20$ (short-dashed blue line), in units of $\ell_B$. The Zeeman couplings 
are $\lambda_{x}=-0.0025$ and $\lambda_z=0.005$. We take 
$B=1$ T and $W=2\ell_B \approx 50$ nm, and 
the energy is measured in units of $E_m \approx 25$ meV.}
\label{P_mono_noex}
\end{figure}
\begin{figure}
 {\includegraphics[width=9cm]{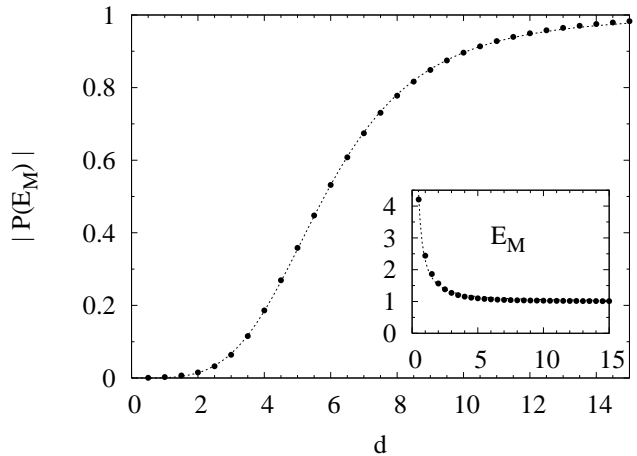}}
\caption{Maximum of the total polarization $|P(E_M)|$ as a 
function of $d$ for the same structure and parameters 
as in Fig.~\ref{P_mono_noex}. 
In the inset: plot of the energy $E_M$ where the peak 
of $|P(E)|$ is located as function of $d$.}
\label{fit}
\end{figure}

\begin{figure}
 {\includegraphics[width=9cm]{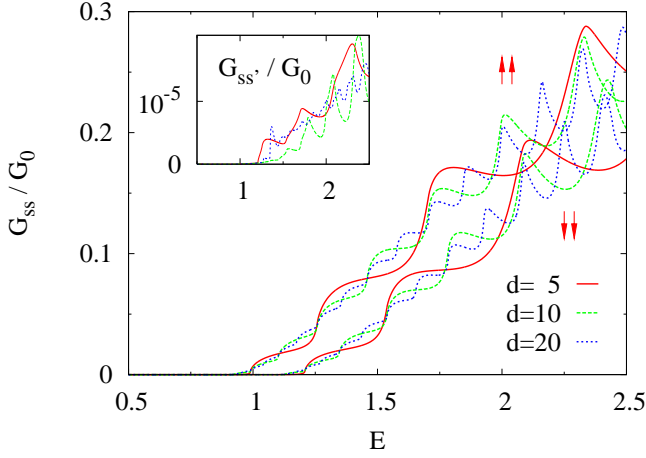}}
\caption{(Color online) Plot of the spin-resolved
conductances $G_{\uparrow\uparrow}$ and $G_{\downarrow\downarrow}$ 
for the same structure as in Fig.~\ref{P_mono_noex}.
Here the Zeeman couplings are 
$\lambda_{x}=-0.2$ and $\lambda_z=0.005$. 
Same line-styles both for spin-up and spin-down. 
In the inset the plot of $G_{\uparrow\downarrow}=G_{\downarrow\uparrow}$ 
for the same parameters. 
}
\label{cond_mono}
\end{figure}

\begin{figure}
 {\includegraphics[width=9cm]{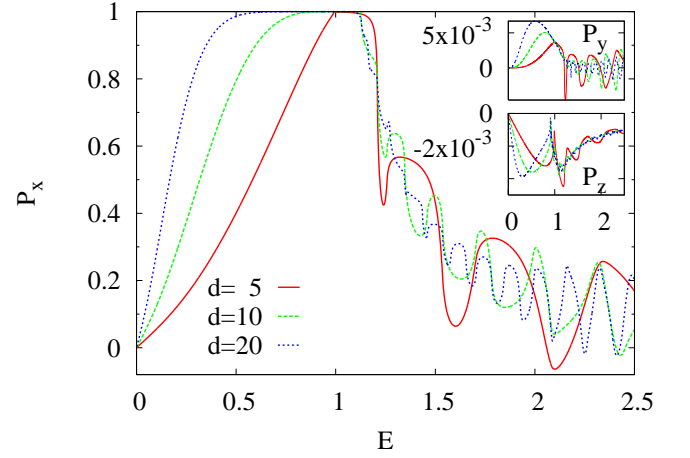}}
\caption{(Color online) 
Plot of the three components of the polarization vector, $P_x$ in 
the main figure, $P_y$ and $P_z$ in the insets, 
for the same structure and parameters as in Fig.~\ref{cond_mono}. 
}
\label{P_mono}
\end{figure}

In this case the wavefunction must be written as
\begin{equation}
\psi = \sum_{\sigma=\pm} \psi_\sigma | \sigma \rangle =
\left( \begin{array}{c}
\psi_\uparrow \\
\psi_\downarrow
\end{array}
\right),
\end{equation}
where $|\sigma=\pm =\uparrow \downarrow\rangle$ 
are the eigenstates of $\sigma_x$ with eigenvalue 
$\sigma=\pm 1$ and $\psi_{\sigma}$ are two-component 
sublattice spinors. 
The solution to Eq. (\ref{DW}) with energy $E$ 
for a state incoming 
from the left with spin projection $\sigma$ 
can be written in the left and right leads as
\begin{eqnarray}
&&\psi(x<0)=\frac{1}{\sqrt{k_x(x)}}{\cal W}(x) \left( \begin{array}{c}
\delta_{\uparrow ,\sigma} \\ r_{\uparrow\sigma} \\
\delta_{\downarrow,\sigma} \\
 r_{\downarrow\sigma} 
\end{array} \right),\\
&&\psi(x>d)=\frac{1}{\sqrt{k_x(x)}}{\cal W}(x) \left( \begin{array}{c}
t_{\uparrow\sigma}\\0\\t_{\downarrow\sigma}\\0
\end{array} \right), 
\end{eqnarray}
where $\delta_{\sigma,\sigma'}$ is the Kronecker delta and
we introduce the $4\times 4$ matrix ${\cal W}$ given by
\begin{widetext}
\begin{equation}
{\cal W}(x) = \left( 
\begin{array}{cccc}
e^{ik_{x\uparrow}(x)x} & e^{-ik_{x\uparrow}(x)x}& 0 & 0\\
e^{i\phi_\uparrow(x)}e^{ik_{x\uparrow}(x)x} & -e^{-i\phi_\uparrow(x)}
e^{-ik_{x\uparrow}(x)x}& 0 & 0\\
0 & 0 & e^{ik_{x\downarrow}(x)x} & e^{-ik_{x\downarrow}(x)x} \\
0 & 0 &e^{i\phi_\downarrow(x)}e^{ik_{x\downarrow}(x)x} & 
-e^{-i\phi_\downarrow(x)}
e^{-ik_{x\downarrow}(x)x}
\end{array}
\right),
\end{equation}
\end{widetext}
with
\begin{eqnarray}
k_{x\sigma}&=& \sqrt{(E-\sigma \lambda_x)^2-(k_y+A)^2},\\
\pm e^{\pm i\phi_\sigma} &=& 
\frac{\pm k_{x\sigma}+i(k_y+A)}{E-\sigma \lambda_x}.
\end{eqnarray}
The transfer matrix is then given by 
\begin{eqnarray}
\hat T = {\cal W}(0^-)^{-1}  \Omega(+)
{\cal W}(0^+){\cal W}(d^-)^{-1} \Omega(-)
{\cal W}(d^+), \nonumber
\end{eqnarray}
where the $4\times 4$ matrix $\Omega(\pm)=e^{\pm i\lambda_z \tau_x \sigma_z}$ 
(non-diagonal in spin space) implements the matching conditions
at the $\delta$ barriers,
and $x^\pm \equiv x\pm 0^+$.
For an incident particle with spin $\sigma$ the continuity 
condition implies
\begin{eqnarray}
\left( \begin{array}{c}
\delta_{\uparrow,\sigma} \\ r_{\uparrow\sigma} \\
\delta_{\downarrow,\sigma} \\
 r_{\downarrow\sigma} 
\end{array} \right)
 = \left( \frac{\cos \phi_i}{\cos\phi_f} \right)^{1/2}\hat T
\left( \begin{array}{c}
t_{\uparrow\sigma}\\0\\t_{\downarrow\sigma}\\0
\end{array} \right),
\end{eqnarray}
from which the transmission amplitudes $t_{\sigma\sigma'}$ are 
easily obtained.
Then using Eqs. (\ref{G})-(\ref{Pz}) 
we can calculate the spin-resolved conductance 
and the spin polarization of the outgoing current 
for an unpolarized incoming current. 
The results are illustrated in Figs. 
\ref{P_mono_noex}-\ref{P_mono}.
The $x$ component of the polarization 
vector $P_x$ is plotted in Fig.~\ref{P_mono_noex} 
as a function of energy for different values of the 
distance $d$ between the two magnetic barriers. 
Just for convenience, here and below we take 
$B^{eff}_x$ negative, so that the spin-up current 
is favored against the spin-down current and $P_x$ 
is mostly positive. 
For the Zeeman couplings we use the values 
$\lambda_x=\mu_B B_x/E_m \approx -0.0025$ and 
$\lambda_z=\lambda_x \frac{B}{B_x}\frac{W}{\ell_B}\approx 0.005$ 
corresponding to  $-B_x=B=1$ T and 
$W= 2\ell_B\approx 50$ nm. 
The spin-resolved conductances 
$G_{\uparrow\uparrow}$ and $G_{\downarrow\downarrow}$ are 
plotted in the inset of Fig.~\ref{P_mono_noex}. 
The conductance $G_{\downarrow\downarrow}$ 
is slightly lower than $G_{\uparrow\uparrow}$ while 
$G_{\uparrow\downarrow}=G_{\downarrow\uparrow}$ are negligibly 
small. 
There is then a very narrow energy region, close to  
$E=E_{th}$ ($E_{th}=1$ for $B=1$ T and $W=2\ell_B$), 
in which $G_{\uparrow\uparrow}$ and 
$G_{\downarrow\downarrow}$ are different (see the inset in 
Fig.~\ref{P_mono_noex}), and $P_x$ exhibits a narrow peak. 
We checked that, with these values of the parameters, 
$P_y$ and $P_z$ are of order $10^{-3}-10^{-4}$, 
thus negligible, and $P_x$ gives the most important 
contribution to the total polarization $|P|=\sqrt{P_x^2+P_y^2+P_z^2}$. 
We have calculated $|P|$ for different values of the distance 
between the barriers $d$ and  found that the polarization maximum 
increases with $d$. This behavior is clearly seen in Fig.~\ref{fit}, 
where the maximum of $|P|$ is plotted as a function of $d$. 
The energy $E_M$ at which the polarization reaches the maximum
is plotted in the inset also as a function of $d$. 
We observe that the height of the polarization peak 
increases with $d$ and its position meanwhile shifts 
towards lower energy.  
From the numerical curves we can extract the following behavior 
for $|P(E_M)|$ and $E_M$:
\begin{eqnarray}
\label{PM}
|P(E_M)|&\simeq& \frac{d^{\alpha}}{C_1+ d^{\alpha}},\\
E_M &\simeq& E_{th}+ \frac{C_2}{d^{\beta}},
\label{EM}
\end{eqnarray}
where $E_{th}=1$, $C_1\simeq 1090$,  $C_2\simeq 1.3$, 
$\alpha\simeq 4.0$ and $\beta\simeq 1.3$. 
In Fig.~ \ref{fit}  the dots represent the exact numerical 
results while the dashed line is the fitting  curve.
Upon increasing $d$ $E_M$ approaches the transmission 
threshold relative to the first barrier. 
From our numerical results we also observe 
that $C_1$ grows 
by decreasing $|\lambda_x|$ approximately as  
$C_1\propto |\lambda_x|^{-1}$, while the other parameters in 
Eqs. (\ref{PM}), (\ref{EM}) only weakly 
depend on the Zeeman couplings. 
For small values of $|\lambda_x|$  we do not have an efficient 
spin filter since the polarization peak occurs in a very narrow 
range of energy where $G_{\uparrow\uparrow}$ and 
$G_{\downarrow\downarrow}$ are both very small. 

In the presence of an exchange field, instead, 
the effective Zeeman interaction also includes the exchange 
contribution and it is thus much larger, $\lambda_x\simeq -0.2$. 
In this case the energy range where we get polarization 
effects is widened and in the spin-resolved conductance plot, 
Fig.~\ref{cond_mono}, we can now clearly distinguish
$G_{\uparrow\uparrow}$ from $G_{\downarrow\downarrow}$.  
In particular, within a range of approximately $10$ meV  
we can have transmission of particles with spin-up and almost 
perfect reflection of particles with spin-down, 
realizing a very efficient spin filter.
Indeed, in Fig.~\ref{P_mono} we can see that, with a ferromagnetic 
region of width $d\sim 20 \ell_B \sim 500$ nm, particles of energy 
$\Phi/2+\lambda_x \lesssim E \lesssim  \Phi/2-\lambda_x$ 
(we set $\lambda_x$ negative), i.e. 
between approximately $20$ and $30$ meV, get perfectly spin-filtered 
upon crossing the magnetic structure.
The polarization, whose largest
component is $P_x$, reaches value one and it is sizable even for a larger 
range of energy. The other two components of the polarization vector, 
$P_y$ and $P_z$, due to 
the spin-flip processes at the barriers and
shown in the insets of Fig.~\ref{P_mono}, remain very small, 
since $\lambda_z$ is small and confined to very narrow 
regions under the two thin barriers.

\section{Bilayer graphene}
\label{secbilayer}

\begin{figure}
{\includegraphics[width=13cm]{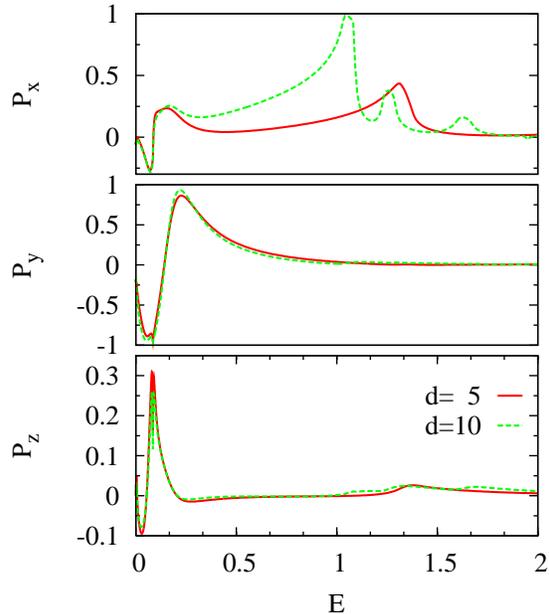}}
\caption
{(Color online) Plot of the polarization vector in bilayer graphene 
for a structure consisting of two $\delta$ barriers 
with opposite signs of the magnetic field, $W=2\ell_B$, at distance 
$d=5$ (solid red line) and $10$ (long-dashed green line), 
in units of $\ell_B$. The dimensionless Zeeman 
couplings are $\lambda_{x}=-0.054$ and 
$\lambda_z=0.108$. 
The energy is given in units of 
$E_b\approx 1$ meV, for $B= 1$ T.}
\label{P_bil_noex}
\end{figure}
\begin{figure}
{\includegraphics[width=9cm]{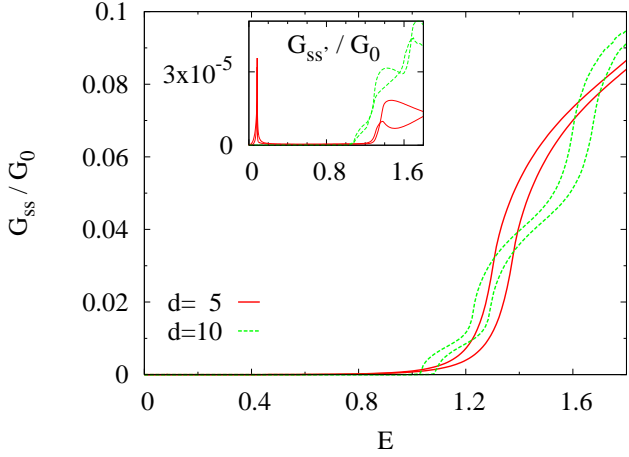}}
\caption
{(Color online) Plot of the spin-resolved
conductances $G_{\uparrow\uparrow}$ and $G_{\downarrow\downarrow}$ 
in bilayer graphene for the same structure and parameters
as in Fig.~\ref{P_bil_noex}.
Same line-styles both for spin-up and spin-down. 
In the inset the plot of $G_{\uparrow\downarrow}$ and 
$G_{\downarrow\uparrow}$ for the same parameters. 
}
\label{cond_bil_noex}
\end{figure}

In this section we consider the spin transport
problem through magnetic structures in bilayer 
graphene. 
In contrast to single-layer graphene, the low-energy dynamics 
of charge carriers in bilayer graphene is governed by a 
quadratic Hamiltonian.\cite{novoselov, falko} Yet, there are 
important differences with respect to a standard 2DEG, 
since the bilayer Hamiltonian is massless and chiral, i.e., 
the wavefunctions for fixed spin projection are two-component 
spinors. 
The effective low-energy Hamiltonian for spinful bilayer graphene 
reads\cite{novoselov, falko,exchangebil}
\begin{equation}
H= \frac{1}{2m}\left( 
\begin{array}{cc}
0 & (\pi_x-i\pi_y)^2 \\
(\pi_x + i\pi_y)^2 & 0
\end{array}
\right) + H_{spin},
\label{bilayer}
\end{equation}
with effective mass $m\approx 0.054 \, m_e$ ($m_e$ is the 
electron mass in vacuum) and
$\pi_{x,y}=p_{x,y}+\frac{e}{c}A_{x,y}$, 
and $H_{spin}$ was defined in Eq. (\ref{hspin}). 
The possibility of inducing an exchange coupling by proximity 
of a ferromagnetic insulator has also been discussed in 
bilayer graphene.\cite{exchangebil} Interestingly, in this case
the electronic band structure can be drastically modified and
a gap may open. However, in the simplest situation, namely  
the bilayer sandwiched
between two ferromagnetic insulators with the same
orientation of the magnetization, the only effect is a spin splitting
and Eq. (\ref{bilayer}) with a large 
in-plane Zeeman coupling is indeed the correct Hamiltonian.\cite{exchangebil}
Again, using $\Psi = e^{ik_yy} \psi$ the problem is reduced to  
a one-dimensional Schr\"odinger equation for a
four-component wavefunction $\psi$.
As in the previous section, all quantities are rescaled 
to be dimensionless, the only difference being that   
the energy scale $E_m=\hbar v_F/\ell_B$ is now 
replaced by $E_b=\hbar^2/2m\ell_B^2$ 
($E_b\approx 1$ meV for $B=1$ T).
Here we directly focus on the most interesting
case of a double resonant barrier configuration 
in the limit of $\delta$ barriers of equal 
and opposite strength 
$\Phi=BW$, with in-plane spin splitting between the barriers.
Several different magnetic field profiles 
have also been studied in Ref. \onlinecite{masir2} but 
only for the spinless case.
With spin quantization axis along the $x$ direction, 
away from the $\delta$ barriers the
elementary solutions of the Schr\"odinger 
equation for spin projection $\sigma$
read
\begin{eqnarray}
U_{\sigma\pm}(x)&=&\left( \begin{array}{c}
1\\
\frac{\left[ \pm k_{x\sigma} + i (k_y+A)\right]^2}{E-\sigma \lambda_x h_{ex,x}}
\end{array} \right)
e^{\pm ik_{x\sigma} x}, \\
V_{\sigma\pm}(x)&=&\left( \begin{array}{c}
1\\
\frac{-\left[\pm q_{x\sigma} - (k_y+A)\right]^2}{E-\sigma \lambda_x h_{ex,x}}
\end{array} \right)
e^{ \pm q_{x\sigma} x} ,
\end{eqnarray}
where 
\begin{eqnarray}
k_{x\sigma}= \sqrt{ (E-\sigma\lambda_x)-(k_y+A)^2}, \\
q_{x\sigma}= \sqrt{ (E-\sigma\lambda_x)+(k_y+A)^2},
\label{kx}
\end{eqnarray}
with $\lambda_x = \mu_B B_x/E_{b}$ ($\lambda_x\approx -0.054$ for 
$B_x=-1$ T) and $\lambda_z= \mu_B BW/ E_b\ell_B$. 
It is convenient to arrange the wavefunction 
$\psi=\psi_\uparrow | \uparrow \rangle + 
\psi_\downarrow |\downarrow \rangle$ 
and its derivative $\psi'$ in a eight-component 
vector and to write it as
\begin{equation}
\left( 
 \psi_\uparrow,
 \psi'_\uparrow,
 \psi_\downarrow,
 \psi'_\downarrow 
\right)^T = {\cal W}(x) A
\end{equation}
where the $8\times 8$ matrix ${\cal W}(x)$ is given by 
\begin{equation}
{\cal W}(x) = 
\left( 
\begin{array}{cccccccc}
U_{\uparrow +} & U_{\uparrow-} & V_{\uparrow+} & V_{\uparrow-} &
0&0&0&0\\
U'_{\uparrow+} & U'_{\uparrow-} & V'_{\uparrow+} & V'_{\uparrow-} &
0&0&0&0\\
0&0&0&0&U_{\downarrow+} & U_{\downarrow-} & V_{\downarrow+} & 
V_{\downarrow-} \\
0&0&0&0&U'_{\downarrow+} & U'_{\downarrow-} & V'_{\downarrow+} & 
V'_{\downarrow-} 
 \end{array}
 \right), \nonumber
\end{equation}
and $A$ is an eight-component vector of complex amplitudes. 
The matching conditions at the positions $x=0,d$ of the
$\delta$ barriers (i.e., continuity of the wavefunction and 
jump of its derivative) can compactly be written as 
\begin{eqnarray}
\left(
\psi_{\uparrow},
\psi'_{\uparrow},
\psi_{\downarrow} ,
\psi'_{\downarrow}
\right)^T(0^-)
 = \Omega(+)
 \left(
\psi_{\uparrow} ,
\psi'_{\uparrow} ,
\psi_{\downarrow},
\psi'_{\downarrow}
\right)^T (0^+),\nonumber \\
\left(
\psi_{\uparrow},
\psi'_{\uparrow},
\psi_{\downarrow} ,
\psi'_{\downarrow}
\right)^T(d^-)
 = \Omega(-)
 \left(
\psi_{\uparrow} ,
\psi'_{\uparrow} ,
\psi_{\downarrow},
\psi'_{\downarrow}
\right)^T (d^+),\nonumber
 \end{eqnarray}
where the matrix $\Omega$, non-diagonal in spin-space, 
is given by
 \begin{equation}  
 \Omega (\pm)=
 \left( \begin{array}{cccc} 
 \tau_0 & 0 & 0 & 0 \\
 \mp  \tau_z & \tau_0  &\mp  \lambda_z \tau_x & 0 \\
 0 & 0 & \tau_0 & 0 \\
\mp  \lambda_z \tau_x & 0 & \mp \tau_z  & \tau_0
 \end{array} \right).
 \end{equation}
Finally, the transfer matrix obtains as
\begin{equation}
\hat T= {\cal W}(0^-)^{-1} \, \Omega(+)\,
{\cal W}(0^+) \, {\cal W}(d^-)^{-1} \, \Omega(-) 
\, {\cal W}(d^+). \nonumber
\end{equation}
The scattering state for a quasiparticle of energy $E> k_y^2$ 
and spin projection $\sigma$ incident on the structure 
from the left can then be written as 
${\cal W}(x)L_\sigma$ for $x<0$ and 
${\cal W}(x)R_\sigma$ for $x>d$, where
\begin{eqnarray}  
&&L^T_\sigma=\left( \delta_{\uparrow,\sigma}, r_{\uparrow\sigma}, 
a_\sigma, 0,
\delta_{\downarrow,\sigma}, r_{\downarrow\sigma}, a'_\sigma, 0 \right), \\
&&R^T_\sigma = 
\left(
t_{\uparrow \sigma},
0 , 0 , b_\sigma , t_{\downarrow \sigma} , 0 , 0 , b'_\sigma
\right) .
\end{eqnarray}
The transmission amplitudes $t_{\sigma\sigma'}$
are found by solving the two linear systems
\begin{equation} 
L_\sigma =\hat T R_\sigma ,
\end{equation}
for $\sigma=\uparrow/\downarrow$.(
This can also be easily generalized to the case of 
unequal strengths of the $\delta$ barriers.)
Then, from Eqs. (\ref{G})-(\ref{P_int}) we can calculate 
the spin resolved conductance and the polarization. 
\begin{figure}
{\includegraphics[width=13cm]{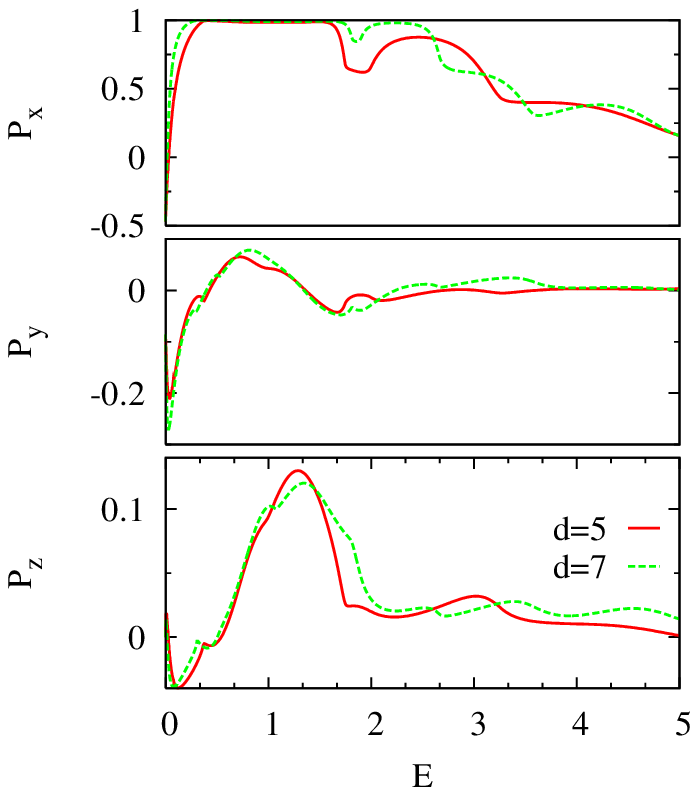}}
\caption
{(Color online) Plot of the polarization in bilayer graphene, 
for a structure consisting of two $\delta$ barriers 
with opposite signs of the magnetic field, $W=2\ell_B$, 
at distance $d=5$ (solid red line) and $7$ (long-dashed green line), 
in units of $\ell_B$. The dimensionless Zeeman couplings
are $\lambda_{x}=-1$ and $\lambda_z=0.108$. 
The energy is given in units of $E_b\approx 1$ meV for $B=1$ T.}
\label{P_bil}
\end{figure}
\begin{figure}
{\includegraphics[width=9cm]{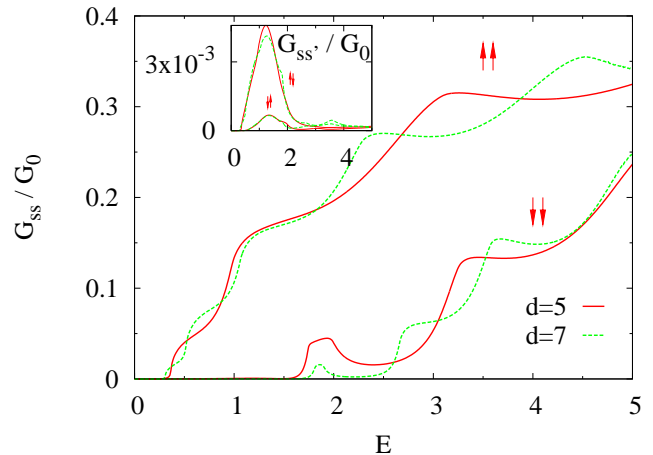}}
\caption
{(Color online) Plot of the spin-resolved
conductances $G_{\uparrow\uparrow}$ and $G_{\downarrow\downarrow}$ 
in bilayer graphene, for the same structure and parameters
as in Fig.~\ref{P_bil}.
Same line-styles both for spin-up and spin-down. 
In the inset the plot of $G_{\uparrow\downarrow}$ and 
$G_{\downarrow\uparrow}$ for the same parameters. 
}
\label{cond_bil_ex}
\end{figure}

In Fig.~\ref{P_bil_noex} all three components of the 
polarization vector are shown for two different values of $d$. 
Looking at the structure of the polarization, we can distinguish 
two different behaviors at two energy scales. The first occurs 
close to $E\sim \lambda_z \simeq 0.1$ and is dominated by 
spin-flip processes, as one can recognize by looking at the 
profile of the three components of the polarization which all 
exhibit some features. Indeed, at this energy scale, $\lambda_z$ 
is not negligible and spin-flips may play a role. At that energy, 
however the conductance is almost zero, see Fig.~\ref{cond_bil_noex}, 
so spin-flips can hardly be detected by direct transport measurements, 
at least for this value of $\lambda_z$. 
The second behavior occurs close to $E\sim E_{th}=\Phi/2=1$ 
which is the signature of a real spin-filter effect, as one can see 
from the spin-resolved conductances plotted in Fig.~\ref{cond_bil_noex}. 
At this energy scale $\lambda_z$ is negligible and,  in fact, 
$P_y$ and $P_z$, as well as
$G_{\uparrow\downarrow}$ and $G_{\downarrow\uparrow}$, 
which are due to spin-flip processes, 
practically vanish, while $P_x$ reaches its maximum.

In the presence of a larger effective in-plane Zeeman coupling 
$\lambda_x$, possibly produced by the exchange spin splitting, 
the spin-filter effect is more pronounced.  
In Fig.~\ref{P_bil} the polarization
 vector is shown for two different value of $d$, with dimensionless 
 Zeeman couplings $\lambda_x=-1$ and $\lambda_z\simeq 0.1$. 
 With this larger absolute value of  $\lambda_x$ the energy range in which the 
 spin filtering occurs is significantly widened.
This is also seen in the behavior of the
spin-resolved conductances in Fig.~\ref{cond_bil_ex}, which 
shows indeed that particles with spin-down are almost perfectly 
reflected by the magnetic structure for energies smaller than 
approximately $2-3$ meV, while spin-up particles are almost always 
transmitted with spin up, since the amplitude for
spin-flips is very small, as 
shown in the inset of Fig.~\ref{cond_bil_ex}.

\section{Conclusions}
\label{conclusions}

In conclusion, we have analyzed the spin 
transport problem through magnetic nanostructures 
in graphene. We have shown that an inhomogeneous 
field profile together with a strong in-plane 
spin splitting can produce a remarkable 
wavevector-dependent spin filtering effect. 
This effect is enhanced 
in a resonant barrier configuration, where the polarization 
can reach values up to one. This result can be understood
by means of a simple kinematical analysis of the problem.

While we confined ourselves to zero temperature, 
we expect that the effect should be observable at
finite temperature as well, as long as the temperature 
is smaller than the in-plane spin splitting. 
If the splitting originates from the exchange
coupling  and the estimate of Ref. {\onlinecite{exchange1}} 
is experimentally confirmed,
then there is a comfortable temperature window (say below 10 K) 
where the effect we discussed could in principle be observed. 
Other mechanisms inducing local ferromagnetic correlations 
in graphene could also be exploited to increase the spin splitting, 
thereby improving the spin filtering effect.
Moreover, since the orbital dynamics and the spin dynamics in 
this problem are to a large extent decoupled 
(this would not be the case with spin-orbit coupling), 
we expect that the addition of a small amount of impurity 
scattering would not spoil the spin filtering effect, 
at least as long as the scatterers are enough long-range
that they do not induce scattering between the two K points.

Along the same lines of this work, one could also investigate
the effects of spin-orbit coupling (SOC). 
While the SOC has so far been estimated 
to be very small in graphene,\cite{so1,so2,so3} recent experimental
results\cite{varykhalov} indicate that in 
quasifreestanding graphene produced on Ni(111) with 
intercalation of Au the Rashba effect leads to 
a large spin splitting of order of 13 meV. 
We plan to address this problem in a forthcoming work.

Finally, we hope that our paper will stimulate further
experimental research on the physics and the transport
properties of magnetic nanostructures in graphene.

\acknowledgments
We gratefully acknowledge R. Egger for
a critical reading of the manuscript.
The work of ADM was supported by the SFB TR 12 of the DFG.

\end{document}